\documentclass[twocolumn,english,superscriptaddress,citeautoscript]{revtex4-2}
\usepackage[T1]{fontenc}
\usepackage[latin9]{inputenc}
\usepackage{array}
\usepackage{multirow}
\usepackage{amsmath}
\usepackage{amssymb}
\usepackage{graphicx}
\usepackage{esint}

\makeatletter

\providecommand{\tabularnewline}{\\}

\usepackage{wasysym}

\providecommand{\tabularnewline}{\\}

\usepackage{babel}
\newcommand{\dbar}{{\mathchar'26\mkern-12mu\mathrm{d}}}

\makeatother

\usepackage{babel}
\begin{document}
\title{Quantum machines using $\mathrm{Cu}_{3}$-like compounds modeled by
Heisenberg antiferromagnetic  in a triangular ring}
\author{Onofre Rojas}
\affiliation{Department of Physics, Institute of Natural Science, Federal University
of Lavras, 37200-900 Lavras-MG, Brazil}
\author{Moises Rojas}
\affiliation{Department of Physics, Institute of Natural Science, Federal University
of Lavras, 37200-900 Lavras-MG, Brazil}
\begin{abstract}
A theoretical study of an antiferromagnetically coupled spin system,
specifically $\ensuremath{\text{Cu}_{3}-\text{X}}(\ensuremath{\text{X=As, Sb}})$,
characterized by a slightly distorted equilateral triangle configuration
is presented. Using the Heisenberg model with exchange and Dzyaloshinskii-Moriya
interactions, g-factors, and an external magnetic field, three quantum
machines are investigated using this system as the working substance,
assuming reversible processes. For $\ensuremath{\text{Cu}_{3}-\text{X}}$
he magnetocaloric effect (MCE) is significant at low temperatures
(around 1K) under a perpendicular magnetic field ($\sim5$T). Although
only the $\ensuremath{\text{Cu}_{3}-\text{As}}$ compound is considered,
since the $\ensuremath{\text{Cu}_{3}-\text{Sb}}$ compound behaves
quite similarly. How MCE influences the Carnot machine, which operates
as a heat engine or refrigerator when varying the external magnetic
field is analyzed. In contrast, the Otto and Stirling machines can
operate as heat engines, refrigerators, heaters, or thermal accelerators,
depending on the magnetic field intensity. The results indicate that
enhanced MCE broadens the operating regions for these machines, with
the Otto and Stirling machines primarily functioning as refrigerators
and accelerators. The corresponding thermal efficiencies are also
discussed for all operating modes.
\end{abstract}
\maketitle

\section{Introduction}

In the past two decades, significant efforts have been made to explore
thermodynamic processes where quantum features of matter are crucial.
This research has successfully combined theoretical proposals \citep{Quan}
with experimental evidence \citep{myers}, bridging the gap between
quantum thermodynamics concepts and practical applications. Quantum
heat engines, like Carnot and Otto, have been studied, with their
characteristics compared to classical counterparts. Additionally,
a quantum heat engine model has been analyzed for efficiency and irreversibility
\citep{kasloff,alicki}. Studies on entangled quantum heat engines
have provided insights into how entanglement influences thermodynamic
efficiency, reinforcing the second law \citep{Zhang,X-He}. Quantum
thermodynamics offers a framework for understanding and optimizing
engines and refrigerators \citep{kasloff-14}. The study of quantum
heat engines based on harmonic oscillators has enhanced our understanding
of optimization and irreversibility \citep{insinga,rezek}. Experiments
with quantum-dot heat engines, which operate without mechanical parts,
have shown near-ideal efficiency, highlighting their relevance to
future quantum technologies \citep{Josefsson}. Moreover, research
on quantum thermal machines in coupled double quantum dots has revealed
various operational regimes \citep{M.Rojas}.

Recent investigations have advanced quantum Carnot, Otto, and Stirling
engines, surpassing previous limitations in achieving Carnot efficiency
with quantum and nanoscale heat engines. These include achieving Carnot
efficiency through semi-local thermal operations with dual bath interactions
\citep{bera}. A six-stroke irreversible quantum Carnot cycle showed
inner friction significantly reduces work output and efficiency \citep{=0000C7akmak20}.
Classical magnetic Otto cycles outperformed quantum ones in work extraction
due to thermodynamic equilibrium \citep{pena}. Energy quantization
effects in the Otto cycle were found to enhance performance, with
experimental proposals using trapped ions \citep{Gelbwaser-Klimovsky}.
The reversible quantum Stirling cycle, involving single or coupled
spin systems, revealed conditions under which a heat engine transforms
into a refrigerator \citep{X.-L.Huang}. Comparisons between quantum
Stirling and Otto cycles for interacting spins highlighted the Stirling
cycle\textquoteright s superiority in work and operational range \citep{=0000C7akmak}.
A quantum Stirling engine based on dinuclear metal complexes offered
tunable modes, with applications in emerging quantum technologies
\citep{C.Cruz}. Several works have explored theoretical dimer systems
as working substances, including quantum engines, refrigerators, and
heaters within a quantum Otto cycle with coupled spins \citep{Oliveira,Makouri,Y-Yin,abd},
and including Dzyaloshinskii--Moriya interactions \citep{asadian}.
Though single-spin and spin lattice systems have been extensively
studied \citep{JZHe,Bender}, trimer systems have been rarely explored,
with one notable quantum Otto cycle analysis \citep{aydinger}.

On the other hand, nanoscale materials with Giant Magnetocaloric Effect
(GMCE) are highly valued for their excellent surface-to-volume ratio,
intensified interactions, and rapid thermal responses, making them
ideal for temperature regulation. Applications include a room-temperature
thermal diode \citep{klinar}, a self-pumping magnetic cooling device
using Mn-Zn ferrite nanoparticles for energy conservation without
external power \citep{chaudhary}, and a ferrofluid-based magnetic
cooling system for efficient heat transfer \citep{Pattanaik}. Additionally,
ferrofluid droplets in microfluidic environments \citep{Sen} and
magnetostructural phase transitions in Ni-Mn-Ga films with notable
MCE at low fields \citep{Qian} have been explored. Other applications
involve energy conversion with gadolinium thick films \citep{ba-becerra,ba-zheng},
magnetic hyperthermia treatments \citep{liu-zhang}, and enhancing
drug delivery via nanocarriers \citep{li-qu}.

The study of magnetic materials is crucial for advancements in spintronics,
nanoscale fabrication, and medical applications. Research on a $\mathrm{Cu}_{3}$
nanomagnet has uncovered phenomena such as half-step magnetization,
hysteresis loops, and asymmetric magnetization, attributed to adiabatic
changes and Dzyaloshinskii-Moriya interactions (DMI) \citep{choi06}.
Spin-electric coupling has also been explored \citep{trif}. Investigations
into $S=1/2$ spin triangle clusters show that their magnetization
and spin configurations are influenced by diamagnetic heteroatoms
($X={\rm As}$ and ${\rm Sb}$) \citep{choi08}, highlighting their
potential in spin-based quantum gates \citep{choi12}. Studies on
spin-frustrated trinuclear copper complexes with triaminoguanidine
revealed strong antiferromagnetic interactions with minimal antisymmetric
exchange \citep{Spielberg}, along with research on various triangular
copper structures \citep{belinsky,Robert,boudalis,stowe,kortz}. $S=1/2$
antiferromagnetic triangular spin rings are promising for studying
unique quantum magnetization effects, such as unusual magnetization
jumps in spin-frustrated $(\ensuremath{\mathrm{VO})_{3}^{6+}}$-triangle-sandwiched
octadecatungstates, which show atypical magnetization steps \citep{yamase}.
The antisymmetric exchange in a tri-copper(II) complex has clarified
its origins and theoretical implications, advancing electronic structure
calculations \citep{Bouammali}.

Theoretical investigations into nanomagnets or magnetic molecular
clusters, which extend beyond experimental observations, are crucial.
Extensive research has documented these insights, with notable theoretical
explorations highlighted in the literature \citep{Szalowski,karlova,torrico-20,torrico-22}.
Recent studies on the $V_{6}$ polyoxovanadate molecular magnet \citep{Szalowski}
used numerical methods to reveal its adaptable magnetocaloric properties.
Another study examined the spin-1/2 Hamiltonian in coupled isosceles
Heisenberg triangles via exact diagonalization, detailing the quantum
phase transition diagram, magnetization at zero temperature, and thermodynamic
behavior \citep{torrico-20}. Further investigation into a $\mathrm{Cu}_{5}$
pentameric molecule using the spin-1/2 Heisenberg model clarified
its thermodynamic characteristics, phase transitions, magnetization,
and magnetocaloric effects \citep{torrico-22}. Karlova et al. explored
antiferromagnetic spin-1/2 XXZ Heisenberg clusters, revealing additional
magnetization plateaus and an enhanced magnetocaloric effect near
magnetization shifts \citep{karlova}.

In our study, we focus on the ${\rm Cu}_{3}-X$ ($\mathrm{X=As,Sb}$)
configuration, with its isosceles triangular structure. Previous research
by Choi et al. \citep{choi06,choi08,choi12} shows that this configuration
fits the Heisenberg model due to its triangular framework. The study
of MCE in ${\rm Cu}_{3}$-like spin systems has clarified their fundamental
properties, particularly at low temperatures \citep{gilberto}. This
paper is structured as follows: Sec. 2 covers the thermodynamics of
the model and quantum machines. Sec. 3 examines the Carnot machine,
Sec. 4 analyzes the Otto machine, and Sec. 5 discusses the Stirling
engine. We conclude with our findings in Sec. 6.

\section{Thermodynamics of the model}

\begin{figure}

\includegraphics[scale=0.4]{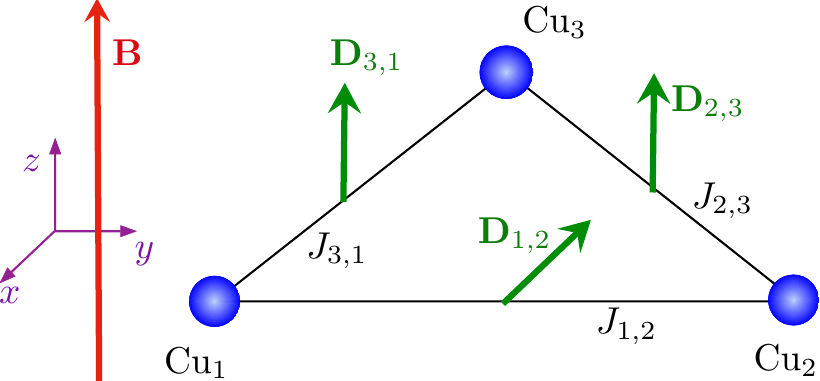}\caption{\label{fig:triang}Schematic representation of the ${\rm Cu}_{3}$-like
system, where the triangle lies in the $xy$-plane.}

\end{figure}

To investigate quantum thermodynamics for machine applications, we
analyze the energy levels of a spin Hamiltonian relevant to ${\rm Cu}_{3}$-like
systems, using the Heisenberg model in an isosceles triangular spin
ring context\citep{choi06,choi08,choi12}. This study builds on previous
work that characterizes the Hamiltonian of ${\rm Cu}_{3}-X$ compounds
\citep{choi06,choi08,choi12}. The Hamiltonian is expressed as
\begin{alignat}{1}
\mathbf{H}= & \sum_{j=1}^{3}\sum_{\alpha=x,y,z}J_{j,j+1}^{\alpha}S_{j}^{\alpha}S_{j+1}^{\alpha}\nonumber \\
 & +\sum_{j=1}^{3}\Bigl[{\bf D}_{j,j+1}\cdot\left({\bf S}_{j}\times{\bf S}_{j+1}\right)+\mu_{B}{\bf S}_{j}\cdot\mathbf{g}_{j}\cdot{\bf B}_{j}\Bigr],\label{eq:H1}
\end{alignat}
where $J_{j,j+1}^{\alpha}$ represents the exchange interaction parameters
between sites $j$ and $j+1$ for $\alpha=\{x,y,z\}$. The vector
${\bf D}_{j,j+1}=(D_{j,j+1}^{x},D_{j,j+1}^{y},D_{j,j+1}^{z})$ represents
the Dzyaloshinskii-Moriya interaction, while $\mathbf{g}_{j}=(g_{j}^{x},g_{j}^{y},g_{j}^{z})$
are the site-dependent g-factors. The magnetic field $\mathbf{B}$
is assumed to be site-independent on the triangle, and $\mu_{B}$
is the Bohr magneton. Parameters were derived from Electron Spin Resonance
(ESR) data \citep{choi06,choi08,choi12}, shown in table \ref{tab:1}.
Only ${\bf D}_{1,2}=(D,D,D)$ is isotropic, while ${\bf D}_{2,3}$
and ${\bf D}_{3,1}$ contribute only to the z-component, ${\bf D}_{2,3}={\bf D}_{3,1}=(0,0,D)$.
A schematic view of the ${\rm Cu}_{3}$-like system is illustrated
in Fig.\ref{fig:triang}.

For convenience, the Hamiltonian \eqref{eq:H1} is expressed in Kelvin
(K) units, redefining $\mu_{B}$ as $\hat{\mu}_{B}=\frac{\mu_{B}}{k_{B}}=0.6717156644$K/T,
where $k_{B}$ denotes the Boltzmann constant, thereby simplifying
the measurements of the magnetic field $\mathbf{B}$ in Tesla (T)
units. This conversion implies all subsequent units are in terms of
$k_{B}$.

\begin{table}
\caption{\label{tab:1}Magnetic parameters of the \{${\rm Cu}_{3}-X$\} compounds
adjusted to experimental data in units of Kelvin, where $X$ denotes
either As or Sb, as extracted from Ref. \citep{choi08}.}
\begin{tabular}{|r|c|c|}
\hline 
Parameters & \{${\rm Cu}_{3}-{\rm As}$\} & \{${\rm Cu}_{3}-{\rm Sb}$\}\tabularnewline
\hline 
\hline 
$J_{1,2}^{x}=J_{1,2}^{y}$ & $4.50$ K & $4.49$ K\tabularnewline
\hline 
$J_{1,2}^{z}$ & $4.56$ K & $4.54$ K\tabularnewline
\hline 
$J_{2,3}^{x}=J_{2,3}^{y}=J_{3,1}^{x}=J_{3,1}^{y}$ & $4.03$ K & $3.91$ K\tabularnewline
\hline 
$J_{2,3}^{z}=J_{3,1}^{z}$ & $4.06$ K & $3.96$ K\tabularnewline
\hline 
$D_{1,2}^{z}=D_{2,3}^{z}=D_{3,1}^{z}$ & $0.529$ K & $0.517$ K\tabularnewline
\hline 
$D_{1,2}^{x}=D_{1,2}^{y}$ & $0.529$ K & $0.517$ K\tabularnewline
\hline 
$g_{1}^{x}=g_{1}^{y}$ & $2.25$ & $2.24$\tabularnewline
\hline 
$g_{2}^{x}=g_{2}^{y}$ & $2.10$ & $2.11$\tabularnewline
\hline 
$g_{3}^{x}=g_{3}^{y}$ & $2.40$ & $2.40$\tabularnewline
\hline 
$g_{1}^{z}=g_{2}^{z}=g_{3}^{z}$ & $2.06$ & $2.07$\tabularnewline
\hline 
\end{tabular}

\end{table}

The eigenvalues of the Hamiltonian \eqref{eq:H1} are obtained through
numerical diagonalization, essential for understanding the energy
spectrum of the system under a fixed magnetic field. 

Therefore, the partition function can symbolically be represented
by 
\begin{equation}
\mathcal{Z}(T,B)={\rm tr}\left({\rm e}^{-\mathbf{H}/T}\right)=\sum_{i=1}^{8}{\rm e}^{-\varepsilon_{i}(B)/T},\label{eq:Zp}
\end{equation}
where the eigenvalues $\varepsilon_{i}(B)$ depend of the Hamiltonian
parameters given in table \ref{tab:1}, external magnetic field $\mathbf{B}$
(in Tesla), and the temperature $T$ (in Kelvin). Although any physical
quantity can be derived from $\mathcal{Z}$, the eigenvalues must
be obtained numerically, limiting analytical solutions. This result,
relevant to quantum machines, highlights the importance of understanding
magnetic and thermodynamic properties at the quantum level, crucial
for advancing quantum technologies.

\subsection{Internal energy, entropy, heat and work}

Here we consider the ${\rm Cu}_{3}$-like compound, which has a finite
number of energy levels given by eq.\eqref{eq:H1}. In order to explore
quantum machine, we need to express the average internal energy as
follows: 
\begin{equation}
U(T,B)=\sum_{i=1}^{8}\varepsilon_{i}p_{i}.
\end{equation}
 here $p_{i}$ is the distribution probability for $i$-th energy
level, which depends implicitly of magnetic field $B$ and temperature
$T$, as well as $\varepsilon_{i}$ depends implicitly of magnetic
field $B$.

In quantum thermodynamics, the first law of thermodynamics is expressed
as
\begin{equation}
{\rm d}U=\dbar Q+\dbar W,
\end{equation}
where $\dbar Q=T{\rm d}\mathcal{S}$, with entropy given by
\begin{equation}
\mathcal{S}\left(T,B\right)=-k_{B}\sum_{i=1}^{8}p_{i}\ln\left(p_{i}\right),
\end{equation}
Therefore, we can express $\dbar Q$ as
\begin{equation}
\dbar Q=\sum_{i=1}^{8}\varepsilon_{i}{\rm d}p_{i},
\end{equation}
whereas $\dbar W$ becomes
\begin{equation}
\dbar W=\sum_{i=1}^{8}p_{i}{\rm d}\varepsilon_{i}.
\end{equation}

\subsection{Quantum reversible process}

Work is related to changes in energy levels $\varepsilon_{i}$ and
results from varying an external parameter like the magnetic field
$B$. In a ${\rm Cu}_{3}$-like system, the quantum isothermal process
involves adjusting the magnetic field, energy gaps, and occupation
probabilities to keep equilibrium with the heat bath. In the quantum
isochoric process, no work is done; heat transfers between temperatures
while occupation probabilities $p_{i}$ and entropy $\mathcal{S}$
adjust to reach thermal equilibrium. The quantum adiabatic process
keeps population distributions constant ($dp_{i}=0$) with no heat
exchange ($\dbar Q=0$), although work can still occur \citep{quan05}.

\subsubsection{Conditions for operational modes of a quantum machine}

Adjusting the cycle parameters can reverse the total work done, in
line with the second law of thermodynamics. In a heat engine, the
system absorbs heat from a hot reservoir at $T_{h}$ ($Q_{in}>0$)
and releases it into a cooler reservoir at $T_{l}$ ($Q_{out}<0$),
converting part of this heat into work ($W_{net}>0$). For a refrigerator,
heat is transferred from the cooler reservoir at $T_{l}$ to the hotter
one at $T_{h}$, with the system absorbing heat from $T_{l}$ ($Q_{out}>0$)
and releasing it at $T_{h}$ ($Q_{in}<0$), requiring more work input
than output ($W_{net}<0$). An accelerator uses work ($W_{net}<0$)
to enhance heat flow from $T_{h}$ to $T_{l}$ ($Q_{in}>0$ and $Q_{out}<0$).
A heater uses work ($W_{net}<0$) to produce heat in both reservoirs,
causing heat release in both ($Q_{in}<0$ and $Q_{out}<0$).

\begin{table}[h]
\caption{\label{tab:engines}Characteristic of work and heat allowed by the
second law of thermodynamics. The signal ($+$) means work done by
the system and heat absorbed; ($-$) means work done on the system
and heat released. }
\begin{tabular}{|l|c|c|c|c|}
\hline 
Operation Mode & $W_{net}$ & $Q_{in}$ & $Q_{out}$ & \multirow{1}{*}{Thermal Efficiency}\tabularnewline
\hline 
\hline 
Heat engine & $+$ & $+$ & $-$ & $\eta=\frac{W_{net}}{Q_{in}}$\tabularnewline
\hline 
Refrigerator & $-$ & $-$ & $+$ & $COP=\frac{Q_{in}}{W_{net}}$\tabularnewline
\hline 
Heater & $-$ & $-$ & $-$ & $COP=\frac{Q_{out}}{W_{net}}$\tabularnewline
\hline 
Accelerator & $-$ & $+$ & $-$ & $COP=\frac{Q_{out}}{W_{net}}$\tabularnewline
\hline 
\end{tabular}
\end{table}

\subsection{Thermal efficiency}

In the heat engine mode, the efficiency of quantum engine, as shown
in Table \ref{tab:engines}, with $\eta<1$. The coefficient of performance
(COP) for other operation modes is also detailed in Table\ref{tab:engines}.
Given that COPs are typically greater than one, we introduce an alternative
expression 
\begin{equation}
\kappa=\frac{COP}{1+COP}.\label{eq:kp}
\end{equation}
 This new metric evaluates COP within a bounded range of $0<\ensuremath{\kappa}<1$,
implying that as COP approaches $0$, $\kappa$ similarly trends toward
$0$, while $\kappa$ approaches $1$ as COP tends to infinity. Additionally,
a $COP=1$ corresponds to $\kappa=0.5$. Consequently, the thermal
efficiency for all operational modes can be defined within the range
of $0$ to $1$, with $0$ indicating the worst performance and $1$
the best. This definition is applicable to the refrigerator, heater,
and accelerator, each described in Table \ref{tab:engines}.

From now on, our discussion will focus on the ${\rm Cu}_{3}-{\rm As}$
compound, as the ${\rm Cu}_{3}-{\rm Sb}$ compound demonstrates similar
characteristics, with parameters values closely comparable to those
of ${\rm Cu}_{3}-{\rm As}$, as shown in Table \ref{tab:1}.

\section{Quantum Carnot machine }

The quantum Carnot machine is a theoretical concept showing the highest
efficiency possible for a heat engine. It's key for understanding
how efficiently energy can be used in quantum systems like the ${\rm Cu}_{3}$-like
compound. This idea is really useful in the field of quantum thermodynamics,
helping us learn about how energy works at the quantum level.

\begin{figure}
\includegraphics[scale=0.6]{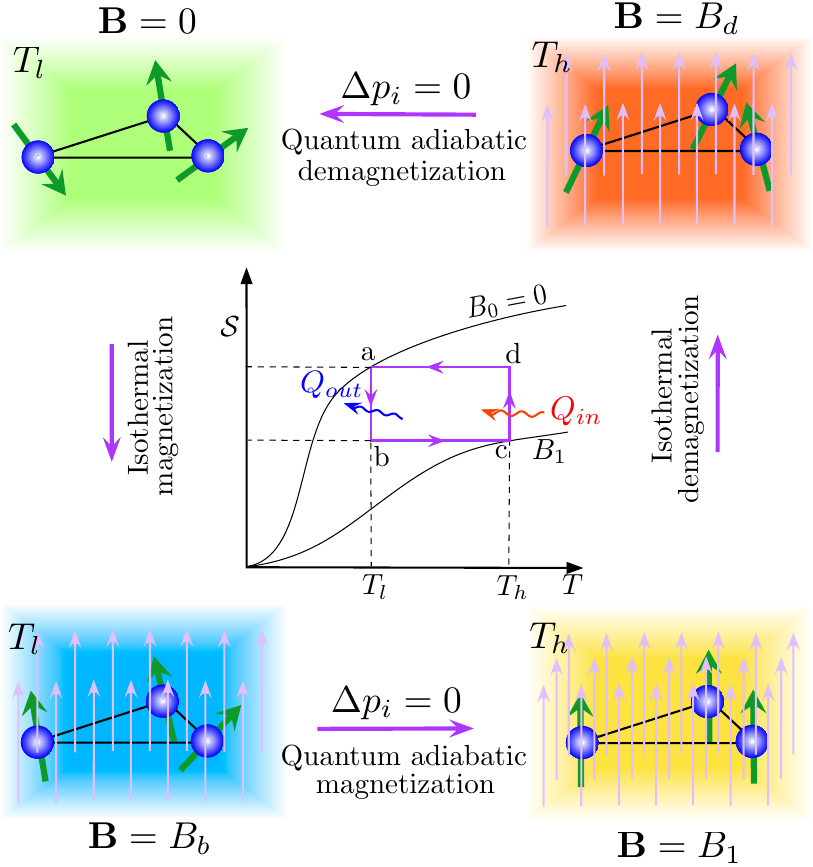}\caption{\label{fig:carnot-cycle}Schematic representation of Quantum Carnot
machine cycle for ${\rm Cu}_{3}$-like compound. Magnetic field is
turned on ($B_{1}$) or turned off ($B_{0}=0$).}
\end{figure}

According to Fig.\ref{fig:carnot-cycle}, we will describe all the
steps of the Carnot cycle below. For simplicity, we will assume, without
losing generality, that when the external magnetic field is increased,
the spins try to align parallel to the external magnetic field. Although
this is not the general rule, frustration can invert this process
(see \citep{gilberto}). However, our analysis will still remain the
same; the only change would be that instead of heat being observed,
heat will be released or vice versa.

\begin{enumerate}
\item Isothermal magnetization process (a-b): The ${\rm Cu}_{3}$-like compound
is placed in a isolated environment. When it comes into contact with
a colder reservoir at temperature $T_{l}$, the external magnetic
field gradually increases, causing the ${\rm Cu}_{3}$-like compound
to release heat to the cold reservoir,
\begin{equation}
Q_{ab}=T_{l}\left[\mathcal{S}(T_{l},B_{b})-\mathcal{S}(T_{l},B_{0})\right],
\end{equation}
 decreasing magnetic entropy and aligning the spins with the magnetic
field. According to the quantum adiabatic condition $p_{i}(T_{l},B_{b})=p_{i}(T_{h},B_{1})$,
which implies that $\mathcal{S}(T_{l},B_{b})=\mathcal{S}(T_{h},B_{1})$,
thus we have 
\begin{equation}
Q_{ab}=T_{l}\left[\mathcal{S}(T_{h},B_{1})-\mathcal{S}(T_{l},B_{0})\right].
\end{equation}
The change in internal energy in the isothermal process at low temperature
$T_{l}$ is 
\begin{equation}
\Delta U_{ab}=U(T_{l},B_{b})-U(T_{l},B_{0}),
\end{equation}
and the corresponding work done is 
\begin{equation}
W_{ab}=Q_{ab}-\Delta U_{ab}.
\end{equation}
\item Adiabatic magnetization process (b-c): The system is again isolated,
and the magnetic field is increased to $B_{1}$ aligning the spins.
Since this process is a quantum adiabatic $Q_{bc}=0$, and the temperature
of the system rises to $T_{h}$. Therefore, the variation energy becomes
\begin{equation}
\Delta U_{bc}=U(T_{h},B_{1})-U(T_{l},B_{b}),
\end{equation}
 and the respective work done is 
\begin{alignat}{1}
W_{bc}= & -\Delta U_{bc}=U(T_{l},B_{b})-U(T_{h},B_{1}),\nonumber \\
= & \sum_{i=1}^{8}\left[\varepsilon_{i}(B_{b})p_{i}(T_{l},B_{b})-\varepsilon_{i}(B_{1})p_{i}(T_{h},B_{1})\right],
\end{alignat}
using the quantum adiabatic condition $p_{i}(T_{l},B_{b})=p_{i}(T_{h},B_{1})$
the work done will become
\begin{alignat}{1}
W_{bc}= & \sum_{i=1}^{8}\left[\varepsilon_{i}(B_{b})-\varepsilon_{i}(B_{1})\right]p_{i}(T_{h},B_{1}),
\end{alignat}
\item Isothermal demagnetization process (c-d). The system is brought into
thermal contact with a hot reservoir at temperature $T_{h}$. The
external magnetic field is gradually decreased. During this process,
the ${\rm Cu}_{3}$-like absorb heat from the reservoir
\begin{equation}
Q_{cd}=T_{h}\left[\mathcal{S}(T_{h},B_{d})-\mathcal{S}(T_{h},B_{1})\right],
\end{equation}
and their spin orientations become more disordered, increasing the
magnetic entropy. Since the quantum adiabatic condition $p_{i}(T_{h},B_{d})=p_{i}(T_{l},B_{0})$,
which implies that $\mathcal{S}(T_{h},B_{d})=\mathcal{S}(T_{l},B_{0})$,
thus we have
\begin{equation}
Q_{cd}=T_{h}\left[\mathcal{S}(T_{l},B_{0})-\mathcal{S}(T_{h},B_{1})\right].
\end{equation}
The variation of internal energy at hot bath with temperature $T_{h}$
is
\begin{equation}
\Delta U_{cd}=U(T_{h},B_{d})-U(T_{h},B_{1})
\end{equation}
The corresponding work done is
\begin{equation}
W_{cd}=Q_{cd}-\Delta U_{cd}.
\end{equation}
\item Adiabatic demagnetization process (d-a). The system is isolated from
any heat reservoir, and the external magnetic field is further decreased.
This process is adiabatic and the temperature of the system drops
to $T_{l}$ without any heat exchange $Q_{da}=0$. And corresponding
variation energy becomes
\begin{equation}
\Delta U_{da}=U(T_{l},B_{0})-U(T_{h},B_{d}),
\end{equation}
 the work done is 
\begin{alignat}{1}
W_{da}= & -\Delta U_{da}=U(T_{h},B_{d})-U(T_{l},B_{0}),\nonumber \\
= & \sum_{i=1}^{8}\left[\varepsilon_{i}(B_{d})p_{i}(T_{h},B_{d})-\varepsilon_{i}(B_{0})p_{i}(T_{l},B_{0})\right],
\end{alignat}
using the quantum adiabatic condition $p_{i}(T_{h},B_{d})=p_{i}(T_{l},B_{0})$
the work done results in
\begin{alignat}{1}
W_{da}= & \sum_{i=1}^{8}\left[\varepsilon_{i}(B_{1})-\varepsilon_{i}(B_{d})\right]p_{i}(T_{l},B_{0}),
\end{alignat}
\end{enumerate}
Finally, the net work is given by
\begin{equation}
W_{net}=W_{ab}+W_{bc}+W_{cd}+W_{da}
\end{equation}
or equivalently
\begin{alignat*}{1}
W_{net}= & Q_{ab}+Q_{cd}.
\end{alignat*}
The heat absorbed and released are respectively $Q_{in}=Q_{ab}$ and
$Q_{out}=Q_{cd}$.

Certainly, it is essential to consider the heat engine efficiency,
denoted as $\eta$ give by $\eta=1+Q_{out}/Q_{in}=1-T_{l}/T_{h}.$
When the Carnot engine operates as a refrigerator, the COP becomes
$COP=Q_{in}/W_{net}=\frac{T_{h}}{T_{h}-T_{l}}.$ Obviously, for the
Carnot engine, both thermal efficiencies are unaffected by the magnetic
field.

\begin{figure}
\includegraphics[scale=0.53]{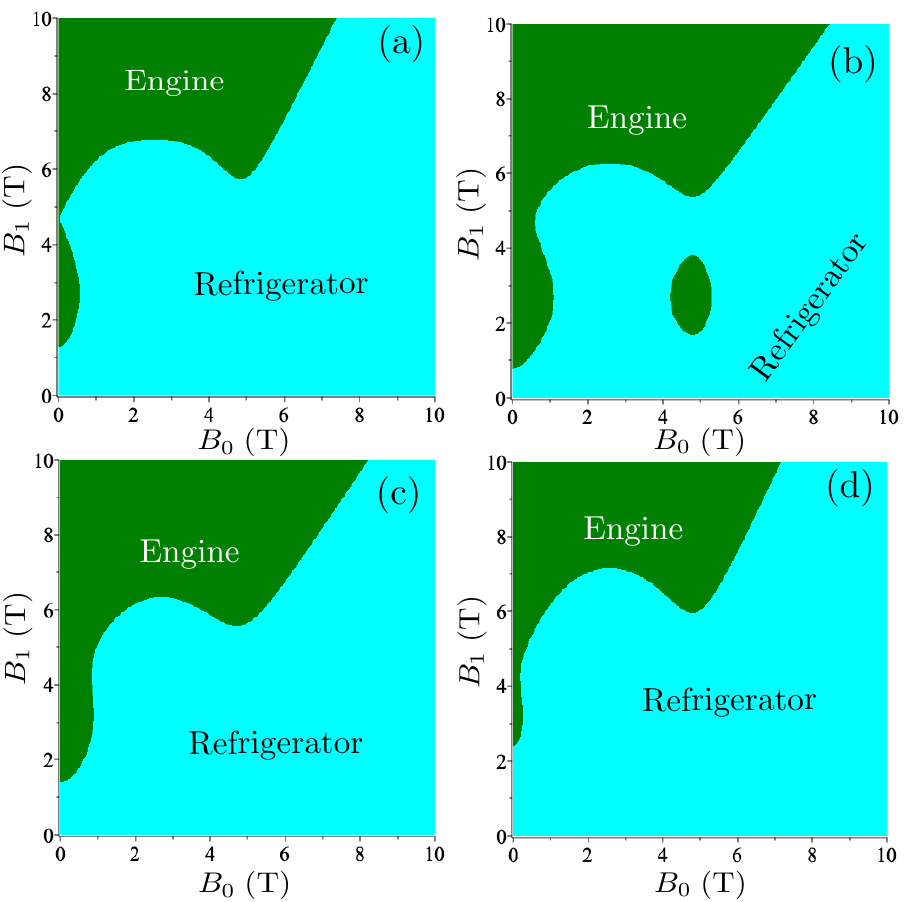}

\caption{\label{fig:Crnt-dgm}Operational modes of the quantum Carnot cycle
for ${\rm Cu}_{3}$-As compound in the $B_{0}$-$B_{1}$ plane. The
green region operates as engine, while the cyan region operates as
refrigerator. (a) For $T_{l}=0.5$K and $T_{h}=1$K. (b) For $T_{l}=0.7$K
and $T_{h}=1$K. (c) For $T_{l}=0.7$K and $T_{h}=1.5$K. (d) For
$T_{l}=1$K and $T_{h}=1.5$K.}
\end{figure}

In Fig. \ref{fig:Crnt-dgm}, we show the operational modes of the
quantum Carnot cycle for a ${\rm Cu}_{3}$-As compound in the $B_{0}$
vs. $B_{1}$ plane. The green region indicates the heat engine mode,
and the cyan region indicates the refrigerator mode. Panel (a) displays
for cold reservoir $T_{l}=0.5$K and hot reservoir $T_{h}=1$K. Assuming
$B_{0}=0$, the system operates as a heat engine for $B_{1}>1.2$
and as a refrigerator at $B_{1}\approx4.7$T. The temperatures and
magnetic fields values were chosen such that the MCE is enhanced\citep{gilberto}.
The heat engine efficiency is $\eta=0.5$ in the green region, with
a refrigerator $COP$ of 2. Panel (b) for $T_{l}=0.7$K and $T_{h}=1$K
shows similar behavior, with refrigeration around $B_{1}\approx4.7$T
and a small heat engine region at $B_{0}\approx5$T and $B_{1}\approx3$T.
The efficiency is $\eta=0.3$, with a COP of 3.333. In panel (c) with
$(\ensuremath{T_{l}=0.7}$K, $\ensuremath{T_{h}=1.5}$K), the small
heat engine region disappears, but operation continues at $B_{0}\approx0T$
and $B_{1}\approx4.7$T. The efficiency is $\eta=0.533$, with a COP
of 1.875. Panel (d) with $T_{l}=1$K and $T_{h}=1.5$K resembles panel
(a), with heat engine operation for $B_{1}>2.5$T, efficiency $\eta=0.333$,
while the COP for the refrigerator is $COP=3$.

\begin{figure}
\includegraphics[scale=0.53]{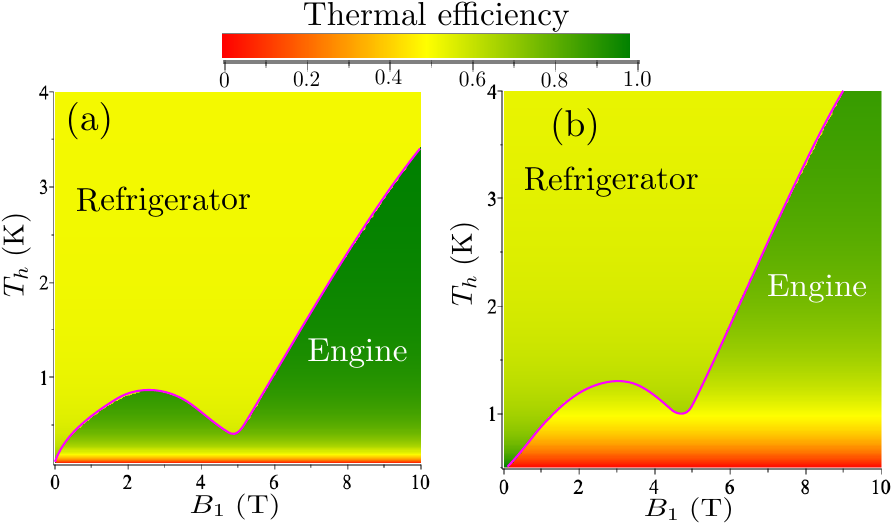}

\caption{\label{fig:Crnt-def}Operational modes of the quantum Carnot cycle
for ${\rm Cu}_{3}$-As compound in the $B_{1}-T_{h}$ plane. Below
the pink curve operates as a heat engine, while above the curve operates
as a refrigerator. (a) For $T_{l}=0.1$K and $B_{0}=0$T. (b) For
$T_{l}=0.5$K and $B_{0}=0$T.}
\end{figure}

Another interesting analysis would involve examining the operational
mode in the $B_{0}-T_{h}$ plane, as illustrated in Fig.\ref{fig:Crnt-def}.
In Panel (a), we consider $T_{l}=0.1$K and $B_{0}=0$. The pink curve
serves as the boundary between the two operational modes: heat engine
and refrigerator. Whereas in panel (c), we present a density plot
that indicates the thermal efficiency under the same conditions as
in panel (a). Here, we observe that the refrigerator operation COP
is consistently above $1$, or equivalently, $\kappa$ is greater
than approximately 0.5. Meanwhile, the efficiency of the Carnot heat
engine remains roughly above $0.8$ in a vast region. As we increase
the temperature $T_{h}$, the system becomes more efficient, leading
roughly $\eta\approx0.96$. Panel (b) shows a similar plot but assumes
a fixed $T_{l}=0.5$K and $B_{0}=0$. Once again, the pink curve delineates
the boundary between the heat engine and refrigerator operation modes.
Essentially, we observe a similar behavior here, although the COP
of the refrigerator consistently exceeds 1, or $\kappa$ is approximately
$0.6$. Meanwhile, the efficiency of the heat engine at low $T_{h}$
is inferior, and for higher temperatures, $\eta$ becomes more efficient
but remains lower than in panel (a). It is worth noting that for each
operational mode, the thermal efficiency remains independent of the
magnetic field $B_{1}$.

Hence, comprehending the quantum Carnot engine of ${\rm Cu}_{3}$-like
compounds is a crucial step in exploring quantum phenomena for practical
technological progress.

\section{Quantum Otto machine}

\begin{figure}
\includegraphics[scale=0.6]{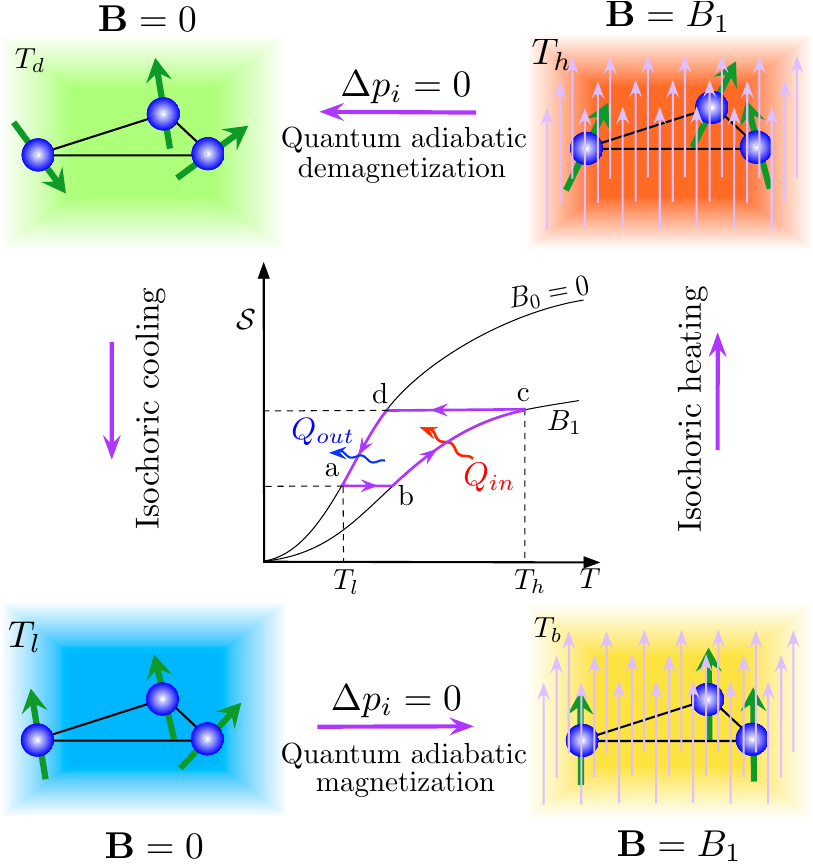}\caption{\label{fig:Otto-cycle}Schematic representation of Quantum Otto machine
cycle for a ${\rm Cu}_{3}$-like compound. Magnetic field is turned
on or turned off.}

\end{figure}

Now we will look at how a compound similar to ${\rm Cu}_{3}$ could
work as a quantum Otto machine as illustrated in Fig.\ref{fig:Otto-cycle}.
This study is important for the field of quantum thermodynamics, as
it offers insights into how ${\rm Cu}_{3}$-like quantum systems might
be used for energy conversion. Understanding the quantum versions
of classical thermodynamic processes can help us better manage energy
at a tiny, microscopic scale.

\begin{enumerate}
\item Adiabatic magnetization process (a-b): In this process, the system
isolates itself from the reservoir. The external magnetic field increases
from $B_{0}$ to $B_{1}$, aligning the spin orientation. No heat
is transferred ($Q_{ab}=0$), even as the temperature rises. Consequently,
the change in energy is
\begin{equation}
\Delta U_{ab}=U(T_{b},B_{1})-U(T_{l},B_{0})
\end{equation}
and the work done is 
\begin{alignat}{1}
W_{ab}= & -\Delta U_{ab}=U(T_{l},B_{0})-U(T_{b},B_{1}),\nonumber \\
= & \sum_{i=1}^{8}\left[\varepsilon_{i}(B_{0})p_{i}(T_{l},B_{0})-\varepsilon_{i}(B_{1})p_{i}(T_{b},B_{1})\right],
\end{alignat}
using the quantum adiabatic condition $p_{i}(T_{b},B_{1})=p_{i}(T_{l},B_{0})$,
we can obtain
\begin{alignat}{1}
W_{ab}= & \sum_{i=1}^{8}\left[\varepsilon_{i}(B_{0})-\varepsilon_{i}(B_{1})\right]p_{i}(T_{l},B_{0}).
\end{alignat}
\item isochoric process (b-c): In this stage, the system comes into thermal
contact with a hot reservoir, and the temperature increases while
keeping the magnetic field constant. No work is done,$W_{bc}=0$.
Consequently, the system absorbs heat, which is given by
\begin{alignat}{1}
Q_{bc}= & \Delta U_{bc}=U(T_{h},B_{1})-U(T_{b},B_{1}),\nonumber \\
= & \sum_{i=1}^{8}\varepsilon_{i}(B_{1})\left[p_{i}(T_{h},B_{1})-p_{i}(T_{b},B_{1})\right].
\end{alignat}
Taking into account the quantum adiabatic condition, $p_{i}(T_{b},B_{1})=p_{i}(T_{l},B_{0})$,
thus the heat becomes
\begin{alignat}{1}
Q_{bc}= & \sum_{i=1}^{8}\varepsilon_{i}(B_{1})\left[p_{i}(T_{h},B_{1})-p_{i}(T_{l},B_{0})\right].
\end{alignat}
\item Adiabatic demagnetization process (c-d): In this stage, the ${\rm Cu}_{3}$-like
system is once again isolated from the reservoir, and the external
magnetic field decreases adiabatically ($Q_{cd}=0$), causing the
system to cool. The corresponding change energy reads
\begin{equation}
\Delta U_{cd}=U(T_{d},B_{0})-U(T_{h},B_{1}),
\end{equation}
 and the work done is .
\begin{alignat}{1}
W_{bc}= & -\Delta U_{bc}=U(T_{h},B_{1})-U(T_{d},B_{0}),\nonumber \\
= & \sum_{i=1}^{8}\left[\varepsilon_{i}(B_{1})p_{i}(T_{h},B_{1})-\varepsilon_{i}(B_{0})p_{i}(T_{d},B_{0})\right].
\end{alignat}
considering the quantum adiabatic condition where $p_{i}(T_{d},B_{0})=p_{i}(T_{h},B_{1})$,
we have
\begin{alignat}{1}
W_{bc}= & \sum_{i=1}^{8}\left[\varepsilon_{i}(B_{1})-\varepsilon_{i}(B_{0})\right]p_{i}(T_{h},B_{1}).
\end{alignat}
\item isochoric process (d-a): In this process, the system comes into contact
with a cold environment, gradually lowering its temperature. Since
there is no change in the magnetic field, no work is done ($W_{da}=0$).
The heat released to the environment is given by
\begin{alignat}{1}
Q_{da}= & \Delta U_{da}=U(T_{l},B_{0})-U(T_{d},B_{0}),\nonumber \\
= & \sum_{i=1}^{8}\varepsilon_{i}(B_{0})\left[p_{i}(T_{l},B_{0})-p_{i}(T_{d},B_{0})\right].
\end{alignat}
Considering the quantum adiabatic condition given by $p_{i}(T_{d},B_{0})=p_{i}(T_{h},B_{1})$,
we obtain
\begin{alignat}{1}
Q_{cd}= & \sum_{i=1}^{8}\varepsilon_{i}(B_{0})\left[p_{i}(T_{l},B_{0})-p_{i}(T_{h},B_{1})\right].
\end{alignat}
\end{enumerate}
Finally, the total work done is given by 
\begin{equation}
W_{net}=\sum_{i=1}^{8}\left[\varepsilon_{i}(B_{1})-\varepsilon_{i}(B_{0})\right]\left[p_{i}(T_{h},B_{1})-p_{i}(T_{l},B_{0})\right].
\end{equation}

\begin{figure}
\includegraphics[scale=0.53]{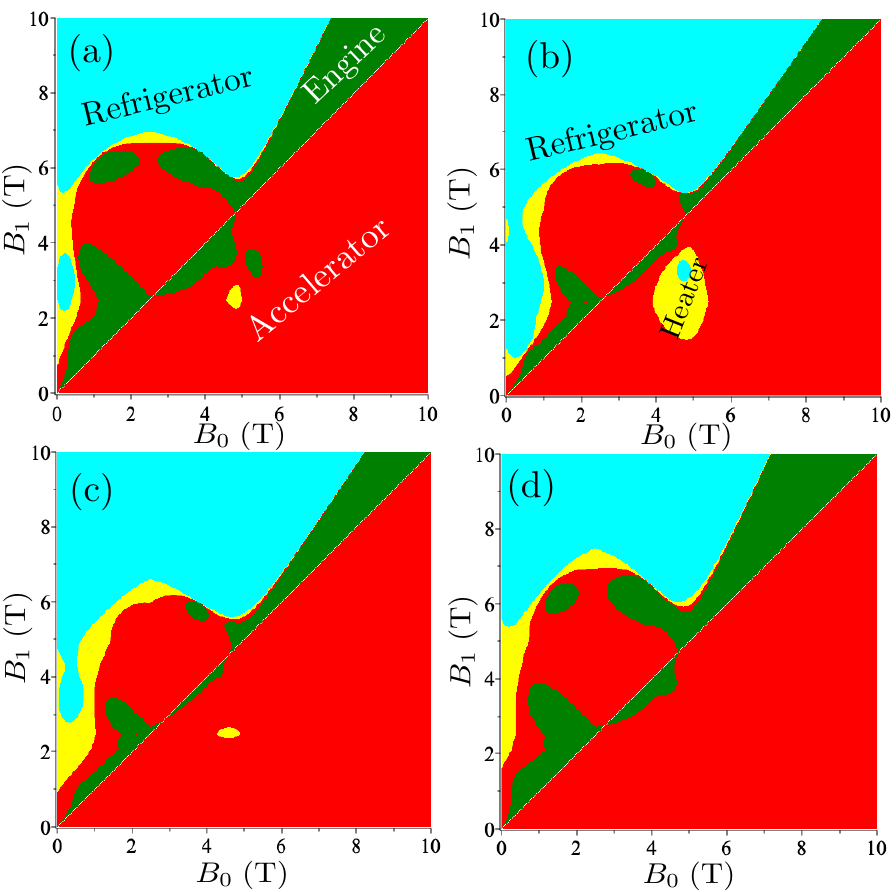}

\caption{\label{fig:Ott-dgm}Operational modes of the quantum Otto cycle for
${\rm Cu}_{3}$-As compound in the $B_{0}-B_{1}$ plane. In green
region operates as an engine, in cyan region operates as refrigerator,
the yellow region operates as heater, and in red region operates accelerator.
(a) For $T_{l}=0.5$K and $T_{h}=1$K. (b) For $T_{l}=0.7$K and $T_{h}=1$K.
(c) For $T_{l}=0.7$K and $T_{h}=1.5$K. (d) For $T_{l}=1$K and $T_{h}=1.5$K.}
\end{figure}

In Fig. \ref{fig:Ott-dgm}, we present the operational modes of the
quantum Otto cycle for a ${\rm Cu}_{3}$-As compound in the $B_{0}-B_{1}$
plane. The color-coded regions signify different modes of operation:
the green region represents the engine mode, the cyan region represents
the refrigerator mode, the yellow region corresponds to the heater
mode, and the red region indicates the accelerator mode. In Panel
(a), we consider the case with $T_{l}=0.5$K and $T_{h}=1$K, the
choice of these parameters is related to the MCE enhancement. Just
to explore the MCE, we assume conveniently $B_{0}$ is approximately
$0.1$T. Notably, the system exhibits peculiar behavior in the range
of $2.2{\rm T}\lesssim B_{1}\lesssim3.8{\rm T}$, where it functions
as a refrigerator. However, at $B_{1}\approx4$T, where the MCE is
significantly enhanced, the system operates as a thermal accelerator.
For $B_{0}>B_{1}$, the system predominantly acts as a thermal accelerator,
while for $B_{1}>B_{0}$, it exhibits four operation modes. Specifically,
for $B_{1}\gtrsim B_{0}$, the system serves as a heat engine, and
where the MCE is enhanced, it behaves as an accelerator. In Panel
(b), with $T_{l}=0.7$K and $T_{h}=1$K, we observe similar behavior.
Although in this case, the ${\rm Cu}_{3}$-As compound could work
as a refrigerator at $B_{1}\approx1$T. However, a small region appears
as a refrigerator at $B_{0}\approx5$T, surrounded by a heater operation
mode. In Panel (c), for $T_{l}=0.7$K and $T_{h}=1.5$K, the region
where the system operates as a refrigerator and heater at $B_{0}\approx5$T
and $B_{1}\approx3$T disappears. However, for $B_{0}\approx0.1$T
and $B_{1}\approx3$T, the system continues to operate as a refrigerator.
Similarly, in Panel (d), for $T_{l}=1$K and $T_{h}=1.5$K, the behavior
is somewhat akin to panel (a), although for $B_{0}=0$ and $B_{1}\gtrsim5$T,
it operates as a refrigerator.

In contrast to the Carnot engine discussed in the previous section,
the Otto engine can operate in four distinct modes: heat engine, refrigerator,
heater, and accelerator.

\begin{figure}
\includegraphics[scale=0.53]{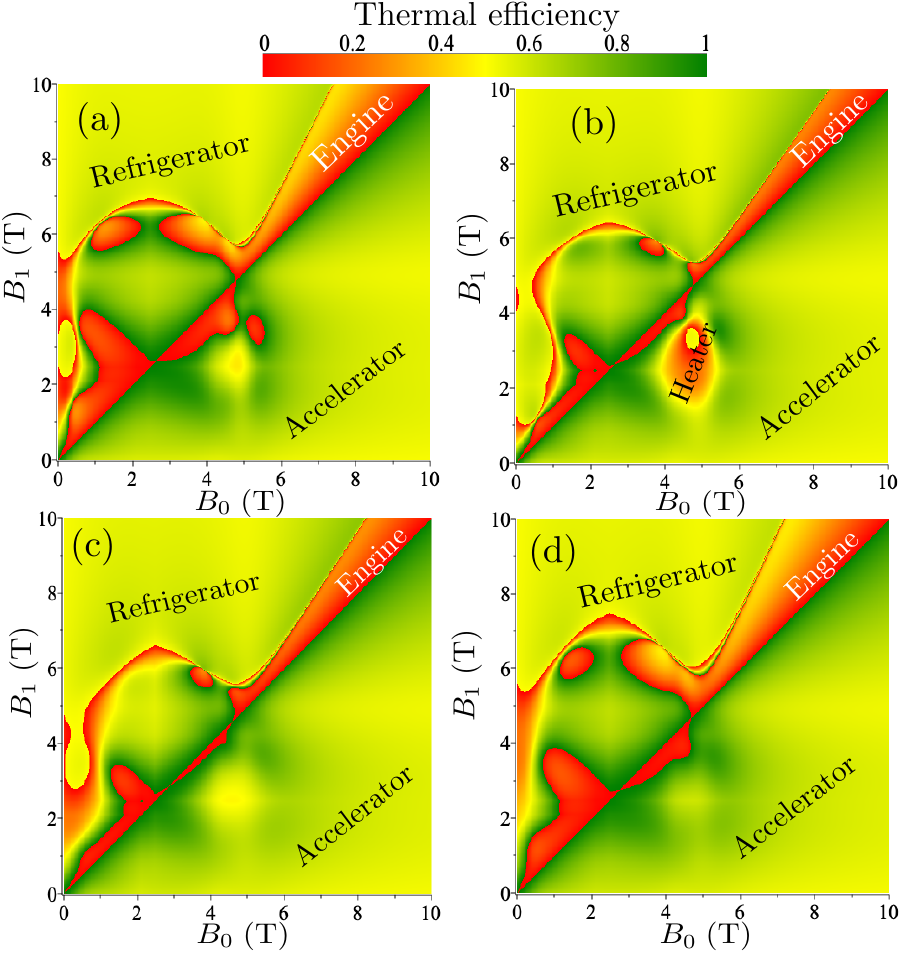}

\caption{\label{fig:Ott-dgm-eff}Thermal efficiency of the quantum Otto cycle
for ${\rm Cu}_{3}$-As compound in the $B_{0}$-$B_{1}$ plane. The
cycle is evaluated under varying cold ($T_{l}$) and hot ($T_{h}$)
temperature conditions: (a) For $T_{l}=0.5$ and $T_{h}=1$. (b) For
$T_{l}=0.7$ and $T_{h}=1$. (c) For $T_{l}=0.7$ and $T_{h}=1.5$.
(d) For $T_{l}=1$ and $T_{h}=1.5$.}
\end{figure}

For engine heat, we use $\eta$ to represent thermal efficiencies,
while for different COPs for the refrigerator, heater, and accelerator,
we employ the eq. \eqref{eq:kp}. This information is presented in
Fig.\ref{fig:Ott-dgm-eff}, using the same parameter set as in Fig.\ref{fig:Crnt-dgm}.
In this representation, regions shaded in red indicate poor efficiency,
green regions correspond to the best efficiency, and yellow regions
indicate efficiency values of $\eta=0.5$. Similarly, for COPs, $\kappa=0.5$
corresponds to $COP=1$. Broadly, we observe that $\kappa\gtrsim0.5$
or $COP\gtrsim1$, while for heat engines, $\eta\lesssim0.5$. This
essentially means that no heat engine achieves an efficiency greater
than $\eta>0.5$. 

\begin{figure}
\includegraphics[scale=0.53]{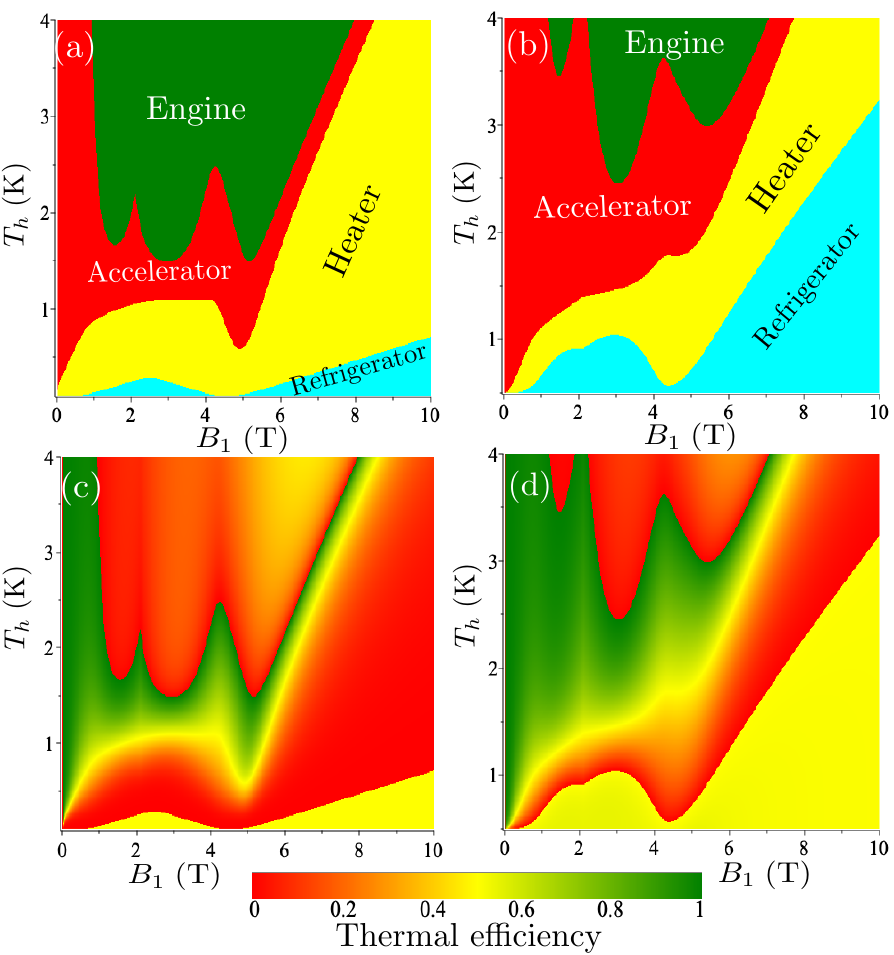}

\caption{\label{fig:Ott-dgm-def}(a) Operation mode of the quantum Otto cycle
for ${\rm Cu}_{3}$-As compound in the plane $B_{1}-T_{h}$, for $T_{l}=0.1$
and $B_{0}=0$. (b) For $T_{l}=0.5$ and $B_{0}=0$. (c-d) Thermal
efficiency under the same condition to (a-b) respectively.}
\end{figure}

Additionally, we can analyze the quantum operational modes as functions
of the hot temperature $T_{h}$ and magnetic field $B_{1}$, assuming
$B_{0}=0$T. In contrast to the Carnot engine, thermal efficiency
now depends on the magnetic fields $B_{0}$ and $B_{1}$, as well
as the temperatures $T_{l}$ and $T_{h}$. In Fig.\ref{fig:Ott-dgm-def},
Panel (a) illustrates the operational modes of the quantum Otto cycle
for a ${\rm Cu}_{3}$-As compound in the $B_{1}-T_{h}$ plane, assuming
fixed values of $T_{l}=0.1$K and $B_{0}=0$. Under these conditions,
we observe all four operational modes: engine (green), refrigerator
(cyan), heater (yellow), and accelerator (red). Panel (b) presents
a similar plot but assumes $T_{l}=0.5$K and $B_{0}=0$T. It demonstrates
that at lower temperatures, the system can function as a refrigerator,
although this region significantly decreases for lower $T_{h}$ values.
Conversely, for higher temperatures, the system tends to operate as
a heat engine. In panels (c-d), we depict the thermal efficiency under
the same conditions as in panels (a-b), respectively. For thermal
efficiency, we use $\eta$ for the heat engine and $\kappa$ for the
refrigerator, heater, and accelerator, as detailed in Table \ref{tab:engines}.
Here, we can observe that thermal efficiency depends on $T_{h}$ and
$B_{1}$. The COP of the refrigerator is roughly around $COP=1$ or
$\kappa=0.5$, while the efficiency of the heat engine is approximately
$\eta\sim0.2$. The COP for the heater is also quite small. However,
the efficiency of the thermal accelerator is notably strong, approaching
$\kappa\rightarrow1$.

Hence, the ${\rm Cu}_{3}$-As compound may offer certain advantages,
mainly in low-temperature region around MCE is relevant, around $B_{1}\sim4.7$T,
and for temperature $T_{h}\gtrsim2$K, the heat engine operates with
high efficiency roughly around $\eta\sim0.5$.

As technology increasingly miniaturizes and ventures into the quantum
domain, insights from the quantum Otto engine could lead to the development
of highly efficient, novel quantum technologies and devices, with
wide-ranging applications from computing to material science. 

\section{Quantum Stirling machine}

\begin{figure}
\includegraphics[scale=0.6]{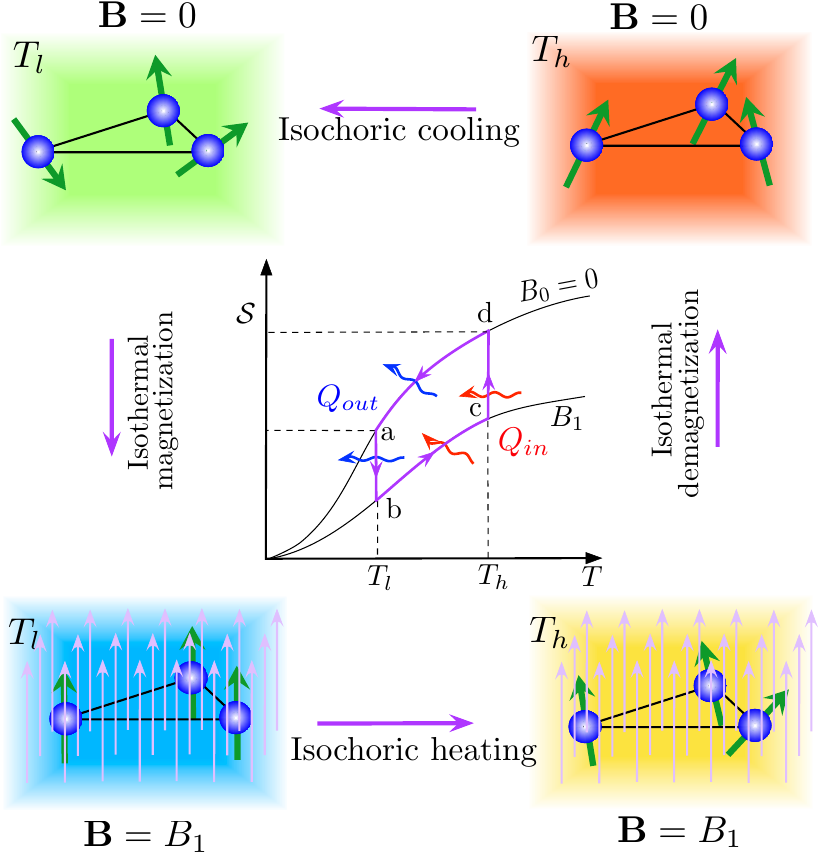}\caption{\label{fig:stirling}Schematic representation of Quantum Stirling
machine cycle for a ${\rm Cu}_{3}$-like compound. Magnetic field
is turned on or turned off.}
\end{figure}

Finally, we consider how ${\rm Cu}_{3}$-like compound could work
as a quantum Stirling machine (see Fig. \ref{fig:stirling})which
represents a relevant stride in the field of quantum thermodynamics,
offering insights into the manipulation and potential applicability
of quantum ${\rm Cu}_{3}$-like systems for energy conversion. By
understanding the quantum mechanical counterparts of classical thermodynamic
processes, could help the energy management at microscopic scales. 

\begin{enumerate}
\item Isothermal magnetization process (a-b). In this process the temperature
is hold constant $T_{l}$, and increases gradually the magnetic field
up to $B_{1}$, consequently the magnetic entropy decreases. The internal
energy variation is given by
\begin{equation}
\Delta U_{ab}=U(T_{l},B_{1})-U(T_{l},B_{0}),
\end{equation}
and the heat released to the environment is 
\begin{equation}
Q_{ab}=T_{l}\left[\mathcal{S}(T_{l},B_{1})-\mathcal{S}(T_{l},B_{0})\right].
\end{equation}
Therefore, the ${\rm Cu}_{3}$-like system work realized in this process
becomes
\begin{equation}
W_{ab}=Q_{ab}-\Delta U_{ab}.
\end{equation}
\item isochoric process (b-c). In this process no work is realized by the
compound, because the magnetic field holds constant $W_{bc}=0$. However,
the temperature of the system increases to $T_{h}$ increasing the
magnetic entropy and spins alignments becoming more erratic, consequently
the system absorb heat from the reservoir, which is given by
\begin{alignat}{1}
Q_{bc}= & \Delta U_{bc}=U(T_{h},B_{1})-U(T_{l},B_{1}),\nonumber \\
= & \sum_{i=1}^{8}\varepsilon_{i}(B_{1})\left[p_{i}(T_{h},B_{1})-p_{i}(T_{l},B_{1})\right].
\end{alignat}
\item Isothermal process (c-d). In this process again the temperature remains
constant at $T_{h}$, while external magnetic field is decreased absorbing
heat from the reservoir 
\begin{equation}
Q_{cd}=T_{h}\left[\mathcal{S}(T_{h},B_{0})-\mathcal{S}(T_{h},B_{1})\right],
\end{equation}
when magnetic field is decreased the spin alignment start disarranging,
leading to higher entropy. Therefore, the work realized in this process
becomes
\begin{equation}
W_{cd}=Q_{cd}-\Delta U_{cd}.
\end{equation}
The variation of internal energy at hot bath with temperature $T_{h}$
is
\begin{equation}
\Delta U_{cd}=U(T_{h},B_{0})-U(T_{h},B_{1}).
\end{equation}
\item isochoric process (d-a). Here the ${\rm Cu}_{3}$-like system is placed
in contact with cold bath consequently decreasing the magnetic entropy
and then the temperature of the system also decreases, but since there
is no change of magnetic field, then no work is realized $W_{da}=0$.
Thus the heat released to the cold bath by the system is given by
\begin{alignat}{1}
Q_{da}= & \Delta U_{da}=U(T_{l},B_{0})-U(T_{h},B_{0}),\nonumber \\
= & \sum_{i=1}^{8}\varepsilon_{i}(B_{0})\left[p_{i}(T_{l},B_{0})-p_{i}(T_{h},B_{0})\right].
\end{alignat}
\end{enumerate}
The total work done can be expressed as a sum of the all works done

\begin{equation}
W_{net}=W_{ab}+W_{cd},
\end{equation}
whereas the heat absorbed by the system becomes

\begin{equation}
Q_{in}=Q_{bc}+Q_{cd}
\end{equation}
and the heat released to the cold bath is 
\begin{equation}
Q_{out}=Q_{da}+Q_{ab}.
\end{equation}

\begin{figure}
\includegraphics[scale=0.53]{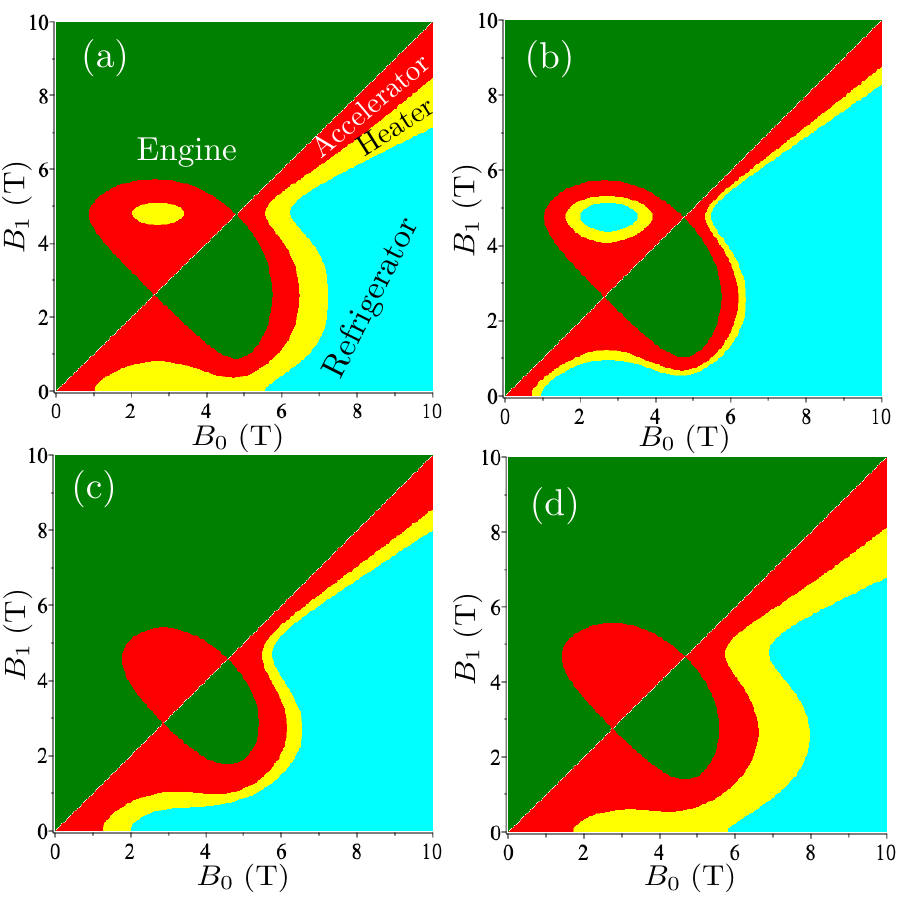}

\caption{\label{fig:str-dgm}Operational modes of the quantum Stirling cycle
for ${\rm Cu}_{3}$-As compound in the $B_{0}$-$B_{1}$ plane. In
green region operates as an engine, in cyan region operates as a refrigerator,
the yellow region operates as a heater, and in red region operates
accelerator. (a) For $T_{l}=0.5$K and $T_{h}=1$K. (b) For $T_{l}=0.7$K
and $T_{h}=1$K. (c) For $T_{l}=0.7$K and $T_{h}=1.5$K. (d) For
$T_{l}=1$K and $T_{h}=1.5$K.}
\end{figure}

In Fig.\ref{fig:str-dgm}, we present the quantum operational modes
of the Stirling cycle for an engine using a ${\rm Cu}_{3}$-As compound
in the $B_{0}-B_{1}$ plane. Panel (a) illustrates this for fixed
temperatures $T_{l}=0.5$K and $T_{h}=1$K. It is clear that when
$B_{1}>B_{0}$, the system mainly operates as a heat engine. Conversely,
when $B_{1}<B_{0}$, the system functions as a refrigerator. Interestingly,
around $B_{0}\sim4.5$T and $B_{1}\sim3$T, there is a region where
it operates as a heat engine. In contrast, around $B_{0}\sim3$T and
$B_{1}\sim4.5$T, the system functions as an accelerator. There is
even a small region at $B_{0}\sim3$T and $B_{1}\sim5$T where it
operates as a heater. Additionally, between these two main operational
modes, there are regions where the heater and thermal accelerator
modes occur, particularly when $B_{1}\lesssim B_{0}$. Panel (b) depicts
the scenario for $T_{l}=0.7$K and $T_{h}=1$K, showing a similar
pattern to panel (a). The main difference is that the heater operation
mode is significantly reduced. However, at $B_{0}\sim3$T and $B_{1}\sim5$,
the system operates as a refrigerator, surrounded by the heater operation
mode. Moreover, for $B_{0}\sim4.5$T and $B_{1}\lesssim1$T, the system
acts as a refrigerator. In Panel (c), we plot the results for $T_{l}=0.7$K
and $T_{h}=1.5$K, which is again similar to (b). However, there are
no heater or refrigerator modes at $B_{0}\sim3$T and $B_{1}\sim5$,
though the accelerator mode still remains at $B_{0}\sim4.5$T and
$B_{1}\lesssim1$T. Finally, Panel (d) presents results for a different
set of parameters, $T_{l}=1$K and $T_{h}=1.5$K. Here, we observe
wider regions for the heater and accelerator modes, but the inversion
operation mode still occurs at $B_{0}\sim3$T and $B_{1}\sim4.5$T,
and vice versa. 

\begin{figure}
\includegraphics[scale=0.53]{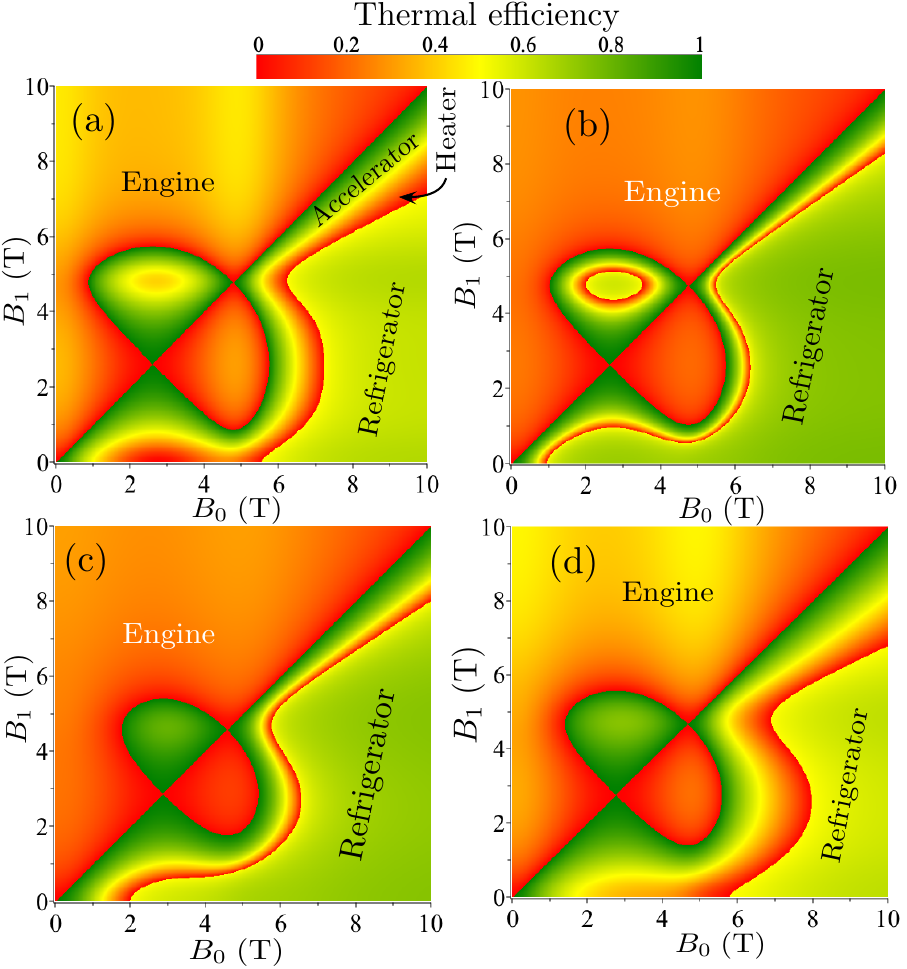}

\caption{\label{fig:str-dgm-eff}Thermal efficiency of the quantum Stirling
cycle for ${\rm Cu}_{3}$-As compound in the $B_{0}$-$B_{1}$ plane.
The cycle is evaluated under varying cold ($T_{l}$) and hot ($T_{h}$)
temperature conditions: (a) For $T_{l}=0.5$ and $T_{h}=1$. (b) For
$T_{l}=0.7$ and $T_{h}=1$. (c) For $T_{l}=0.7$ and $T_{h}=1.5$.
(d) For $T_{l}=1$ and $T_{h}=1.5$.}
\end{figure}

Our next analysis focuses on the thermal efficiency of the operating
modes of the quantum Stirling engine. To conveniently present this,
we display it in Fig.\ref{fig:str-dgm-eff}, assuming the conditions
considered in Fig.\ref{fig:str-dgm}. We calculate the thermal efficiency
using the values provided in Table \ref{tab:engines} and Eq.\eqref{eq:kp}.
When the Stirling machine operates as a heat engine, the efficiency
is notably $\eta\lesssim0.5$ in the vast region where it functions.
This efficiency is even higher in panel (a) for $B_{1}\gtrsim5$T
and when $B_{1}>B_{0}$. On the other hand, when it operates as a
refrigerator, the thermal efficiency is slightly above $\kappa\gtrsim0.5$
or $COP\gtrsim1$, showing better efficiency in panels (b-c). When
the system works as a heater, the thermal efficiency is clearly below
$\kappa<0.5$ in most of the region where it operates. Finally, when
the system functions as a thermal accelerator, the corresponding efficiency
becomes considerably high.

It is worth noting that in the region where the magnetocaloric effect
is more prominent, the quantum operation mode undergoes a type of
inversion. Similar to the Otto machine, the Stirling machine has four
operational modes: heat engine, refrigerator, heater, and accelerator.
The quantum Stirling engine could serve as a prototype for elucidating
the fundamental principles of quantum heat engines and potential technological
application\citep{Das}.

\section{Conclusion}

This paper investigates the ${\rm Cu}_{3}-X$ ($\mathrm{X=As,Sb}$)
antiferromagnetic spin system with a slightly distorted equilateral
triangular configuration. Using the Heisenberg model on a triangular
lattice, it incorporates exchange interactions, Dzyaloshinskii-Moriya
interaction, $g$-factors, and external magnetic fields to represent
system properties as highlighted in prior research \citep{choi06,choi08,choi12}.
The significance of the ${\rm Cu}_{3}$-like system, driven by its
fundamental properties, has grown due to potential applications in
spintronics, nanotechnology, and biomedicine. The MCE in ${\rm Cu}_{3}-X$,
especially at low temperatures (around $T\approx1$K) under perpendicular
magnetic fields ($\sim5$T), is also analyzed \citep{gilberto}

The MCE impact on quantum machine operation and efficiency is significant.
We explored the Carnot, Otto, and Stirling machines within a ${\rm Cu}_{3}$-like
system as the working substance, focusing on how external magnetic
fields and temperature affect their performance assuming reversible
process. The Carnot machine operates as a heat engine for $B_{1}\lesssim B_{0}$
and as a refrigerator for $B_{1}\gtrsim B_{0}$, with a shift in mode
near the MCE. The Otto machine is versatile, mainly functioning as
a thermal accelerator for $B_{1}<B_{0}$ and as a refrigerator for
$B_{1}>B_{0}$, with a small region where it acts as a heat engine
and heater. The Stirling machine alternates between a heat engine
and thermal accelerator near the MCE, operating primarily as a heat
engine for $B_{1}>B_{0}$ and as a refrigerator for $B_{1}<B_{0}$.
We also evaluated their thermal efficiencies for each machines. This
study enhances the understanding of quantum heat machines, especially
in low-temperature regimes where the MCE is crucial, offering insights
into practical applications of the ${\rm Cu}_{3}-X$ system in quantum
thermodynamics. Although our analysis focused solely on the compound
${\rm Cu}_{3}$-As, the second compound ${\rm Cu}_{3}$-Sb will also
exhibit similar behavior for Carnot, Otto, and Stirling machines,
with no relevant changes worth commenting on, as analyzed in Ref.\citep{gilberto}.
We did not include any plots here to avoid repetitive behavior.

Future research could focus on optimizing quantum machines using ${\rm Cu}_{3}$-like
systems through material tuning, analyzing non-ideal quantum conditions,
and expanding applications in quantum refrigeration, assuming non-reversible
quantum machines.
\begin{acknowledgments}
O. R. and M. Rojas thank CNPq and FAPEMIG for partial financial support.
M. Rojas acknowledges FAPEMIG Grant APQ-02226-22.
\end{acknowledgments}

\end{document}